\documentclass{article}
\usepackage[utf8]{inputenc}
\usepackage{amsmath,amssymb,amsfonts}
\usepackage[left=2cm,right=3cm]{geometry}
\usepackage[linesnumbered,ruled]{algorithm2e}
\usepackage{graphicx}
  \usepackage{authblk}
\usepackage{subfig}
\usepackage{float}
\usepackage{xcolor}

\begin{document}

\title{All roads lead to Rome:\\ Many ways to double spend your cryptocurrency
}

\author{Zhiniang Peng}
\author{Yuki Chen}

\affil{Qihoo 360}
\affil{Email : jiushigujiu@gmail.com}

  \maketitle

%

\begin{abstract}
{In 2008, Satoshi Nakamoto proposed an electronic cash system (bitcoin) that is completely realized by peer-to-peer technology. The core value of this scheme is that it proposes a solution based on Proof-of-Work, so that the cash system can run in a peer-to-peer environment and be able to prevent double-spend attacks. Bitcoin has been developed for ten years, and since then countless digital currencies have been created. But the discussion of double-spend attacks seems to still concentrate on 51\% Attacks. In fact, our research has found that there are many other way to achieve double-spend attacks. In this paper, by introducing a number of double-spend attack vulnerabilities that we have found in EOS, NEO and other large blockchain platforms, we summarized various reasons for causing double-spend attacks, and propose an efficient mitigation measure against them.}
\end{abstract}

\section{Introduction}
In 2008, Satoshi Nakamoto proposed an electronic cash system that was completely implemented through peer-to-peer technology, which enabled online payments to be initiated directly by one party and paid to the other without the need to go through any third-party financial institution~\cite{bitcoin}. Although digital signature partially solves this problem, if third-party’s support is still needed to prevent double spending, then the system loses its value. The essence of Bitcoin's Proof-of-Work (PoW) is to make the cash system operate in a peer-to-peer environment and prevent double-spend attacks.

The principle of the PoW mechanism is as follows: each block in the network contains the transaction in the current network as well as the block header hash of the previous block. When a new block is generated, its block header hash must satisfy the PoW requirements(a large number of hash calculations is required). The entire network connects the blocks that satisfy the PoW requirements to generate a blockchain~\cite{blockchain}. Unless the attacker completes the PoW hash calculations all over again, the transaction record will not be changeable. The longest blockchain will not only serve as proof of the observed transaction sequence, but also as a consensus from the majority of the network. As long as most of the computing power in the entire network does not intend to cooperate to attack the network, then honest nodes will generate the longest chain that exceeds the attacker’s, thus preventing double-spend attacks.

A double-spend attack is actually a result. If attacker A pays the same bitcoin to user B and user C, and both users approve the transaction, then we can say that A spends the bitcoin twice and realizes a double-spend attack. Of all the double-spend attacks against PoW mechanism, 51\% attack~\cite{51} is the most heatedly discussed. But in fact, there are a lot of forms of double-spend attacks against PoW, including Finney attack~\cite{Finney}, race condition attack~\cite{bicoinsecurity}, Vector76~\cite{Vector76} attack and so on. These attacks have actually received quite a lot of attention and discussion. However, there are many other forms of practical digital currency double-spend attacks which did not attract people's attention.

In this paper, by introducing a number of double-spend attack vulnerabilities that we have found in EOS~\cite{EOS}, NEO~\cite{NEO} and other large blockchain platforms, we summarized various reasons for causing double-spend attacks, and propose an efficient mitigation measure against them.

\section{Categories of double-spend attack}
Smart contract platform is essentially to share one ledger across the entire network. This can be considered as a distributed state machine replication problem. The current ledger status can be considered as $State_n$. When a new transaction $Tx_{n+1}$ is generated, $Tx_{n+1}$ will have an effect on $State_{n}$, thus $State_{n}$ transits to $State_{n+1}$. This process can be illustrated by formula:
\[State_n \times Tx_{n+1} \rightarrow State_{n+1}\]

The smart contract platform consensus mechanism essentially applies all transactions [$Tx_1$, $Tx_2$ ,..., $Tx_n$] to the initial $State_0$ in order, so that the entire network remains in the same state. Each block in the blockchain actually splits the transaction sequence [$Tx_1$, $Tx_2$,..., $Tx_n$] into different blocks Block1 [$Tx_1$, $Tx_2$], Block2 [$Tx_3$, $Tx_4$] and links them in order. In the process of state machine replicating of the whole network, if the state of the whole network is inconsistent for some reason, we can consider that the whole network generates a fork. The fork can be exploited by the attacker to conduct a double-spend attack.
In this article, we divide these double-spend attack vulnerabilities we have found into three categories:
     \begin{enumerate}
     \item Double-spend attack caused by insufficient verification.
     \item Double-spend attack caused by inconsistent execution of state machine.
     \item Double-spend attack due to consensus mechanism.
	\end{enumerate}

The main reason for double-spend attack caused by insufficient verification exists in the implementation of the block and transaction verification. The bitcoin vulnerability CVE-2018-17144 is such a vulnerability.

Double-spend attack caused by the inconsistent execution of the state machine is mainly due to the inconsistency execution of the smart contract virtual machine for various reasons, thus creating a fork in the entire network and further causing a double-spend attack.

A consensus mechanism vulnerability can create a fork in the entire network, further creating a double-spend attack. The 51\% attack that people frequently talk about is actually a fork vulnerability in the PoW consensus mechanism.

\section{Double-Spending Attack Caused by Insufficient Verification}
The main reason for double-spend attack caused by insufficient verification exists in the implementation of the block and transaction verification. Here we will introduce two vulnerabilities of this kind.

In a blockchain project, the way in which a transaction $Tx_1$ is included in $Block_1$ is always as follows: first, calculate the hash value $Hash_1$ of the transaction $Tx_1$, and then use $Hash_1$ to combine with other transactions' hash values $Hash_2$,...,$Hash_n$ to form Merkle Hash Tree. Calculate the root of the hash tree, and then include the root into $Block_1$. This creates a bond between the transaction and the block. In general, unless an attacker can break the collision resistance of a hash function, he cannot break the binding of a transaction to a block. If the attacker can break the binding, the attacker can achieve a double-spend attack by causing the fork of the whole network. Below we will provide two double-spend attack vulnerabilities we found on NEO, in which attacker can break the binding of transaction and block.

\subsection{GetInvocationScript double-spend vulnerability of NEO VM}
In a blockchain project, a transaction is generally made up of an unsigned part (UnsignedTx, the content to be executed by the transaction) and a signed part (the transaction’s witness). In a blockchain project such as Bitcoin, the hash calculation of the transaction contains the signed part of the transaction. In other various blockchain platforms such as NEO and ONT, the calculation formula of the transaction is hash=SHA256 (UnsignedTx). That is to say, the hash of the transaction is calculated from the unsigned part, regardless of the transaction's witness. When the NEO smart contract is executed, the witness of the transaction can be obtained through the Transaction\_GetWitnesses method. Its specific implementation is in Figure~\ref{NEO1}:
\begin{figure}
\centering
  \includegraphics[scale=0.5]{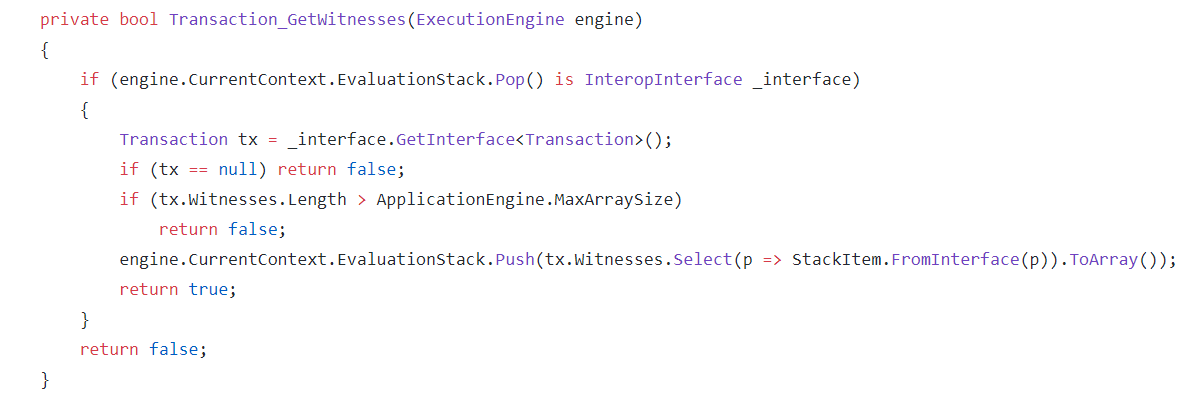}
    \caption{Transaction\_GetWitnesses in NEO}
    \label{NEO1}
\end{figure}

\begin{figure}
\centering
  \includegraphics[scale=0.5]{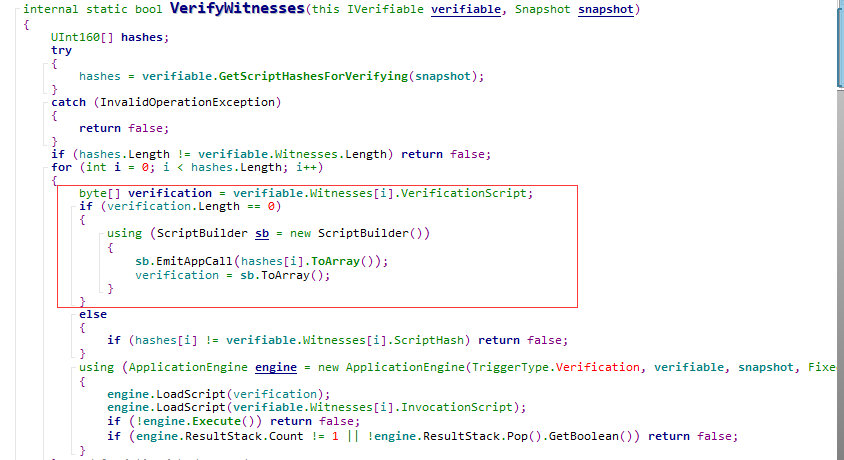}
    \caption{VerifyWitness in NEO}
    \label{NEO2}
\end{figure}

After a contract transaction gets its own witness, it can also get the verification script in the witness through the Witness\_GetVerificationScript method. If the attacker can construct two different verification scripts for the same unsigned transaction UnsignedTx1, he can cause inconsistencies in the execution of the contract. Under normal circumstances, the VerificationScript of the contract is determined by information such as the input of the contract. The attacker cannot construct a different script to pass the verification. However, we found that in the VerifyWitness method in Figure~\ref{NEO2}, when VerificationScript.length=0, the system will call EmitAppCall to execute the target script hash.
So when VerificationScript=0, or VerificationScript is equal to the target script, the witness validation condition can be met. That is to say, an attacker can construct two different VerificationScripts for the same unsigned transaction UnsignedTx\_1. The attacker can take advantage of this to perform double-spend attacks against all token assets on the NEO smart contract. The specific attack scenarios are as follows:
     \begin{enumerate}
     \item The attacker constructs a smart contract transaction $Tx_1$ (unsigned content $UnsignedTx_1$, verification script is $VerficationScript_1$). In the contract execution of UnsignedTx\_1, the contract checks whether its VerficationScript is VerficationScript\_1. If it is VerficationScript\_1, then send the token to user A. If the VerficationScript is empty, the token is sent to user B.
     \item $Tx_1$ is included into block $Block_1$.
     \item After the attacker receives $Block_1$, he replaces $Tx_1$ with $Tx_2$ ($Tx_1$ has the same unsigned content $UnsignedTx_1$ as $Tx_1$, but the verification script is empty) to generate $Block_2$. The attacker sends $Block_1$ to user A and $Block_2$ to user B.
     \item When user A receives $Block_1$, he finds that he has received the token sent by the attacker. When B user receives $Block_2$, he will also find that he has received the token sent by the attacker. The double-spend attack is completed.
	\end{enumerate}
It can be seen that it is very easy to exploit this vulnerability and the exploitation can be used on all tokens on NEO smart contract to double spend the asset. Hence, this vulnerability is very high-risk.

\subsection{Double-spend attack caused by bypassing NEO MerlkeTree Binding}
The binding of smart contract transactions to blocks is usually done through MerkleTree. If the attacker can bypass the MerkleTree binding, he can cause double spending on any transaction. Here let’s have a look at the implementation of NEO's MerkleTree in Figure~\ref{NEO3}:
\begin{figure}
\centering
  \includegraphics[scale=0.3]{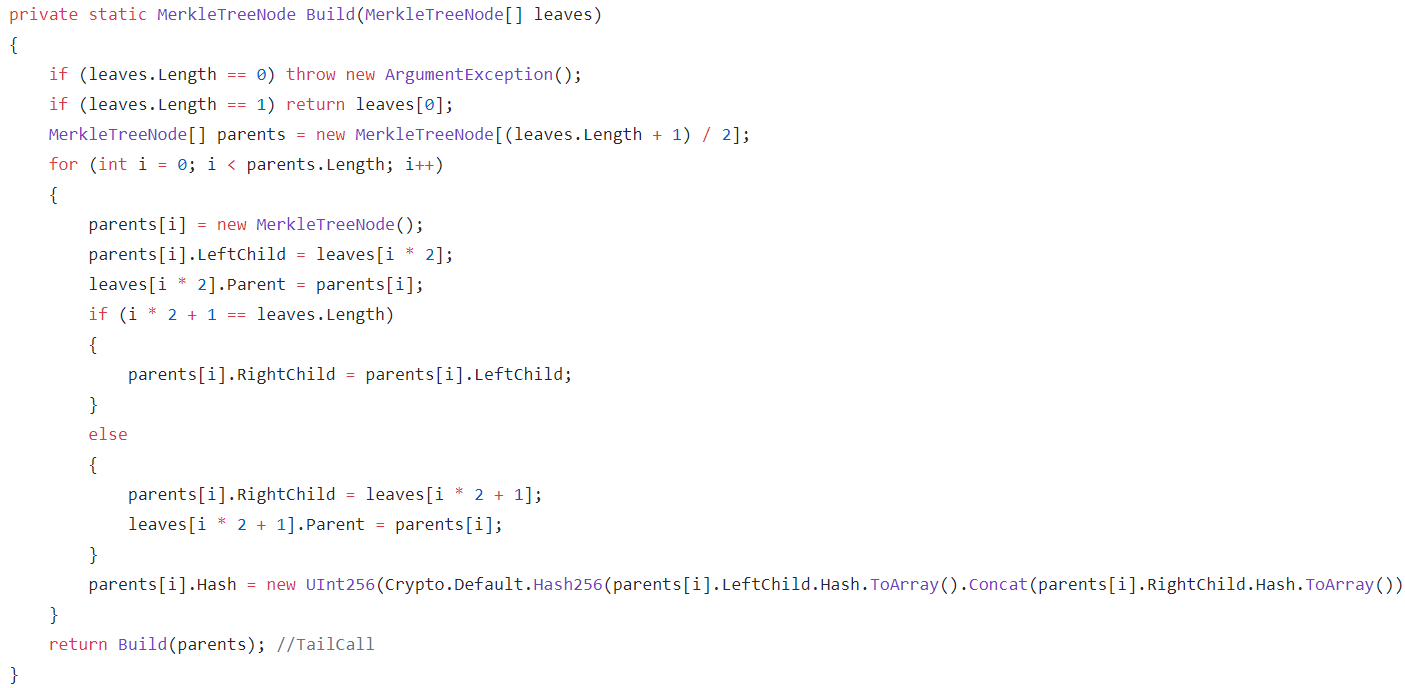}
    \caption{MerkleTree computation in NEO}
    \label{NEO3}
\end{figure}
In the MerkleTreeNode function, NEO performs the calculation from MerkleTree leaf node to the parent node. But there is a problem, that is, when leaves.length is odd n, NEO's MerkleTree will copy the last leaf node once and add it to the calculation of MerkleTree. That is to say, when n is an odd number, the MerkleRoot values of the following two transaction sequences will be equal:
	\[[Tx_1 ,Tx_2 ,\cdots ,Tx_n]\]
	\[[Tx_1 ,Tx_2 ,\cdots ,Tx_n ,Tx_{n+1}]\]
    \[Tx_{n+1}= Tx_n\]

With this feature, attacker can achieve a double-spend attack on any NEO asset. The specific attack scenario is as follows:
     \begin{enumerate}
     \item Assume there is a normal legitimate $Block_1$ and the list of transactions included is [$Tx_1$, $Tx_2$, ... $Tx_n$]. After the attacker receives $Block_1$, he replaces the transaction list with [$Tx_1$, $Tx_2$, ..., $Tx_n$, $Tx_{n+1}$] to generate $Block_2$. Then publish $Block_2$ to the network.
     \item After a common node receives $Block_2$, it will check the validity of $Block_2$. However, because [$Tx_1$, $Tx_2$, ... $Tx_n$, $Tx_{n+1}$] has the same MerkleRoot as [$Tx_1$, $Tx_2$, ..., $Tx_n$]. Therefore, $Block_2$ can pass the block validity check, thus entering the block persistence process. NEO locally cancels the verification of transactions in the legitimate block by ordinary node (trusting several consensus nodes). Then the $Tx_n$ transaction can be executed twice by the ordinary  node – which means the double-spend attack is executed successfully.
	\end{enumerate}
It can be seen that it’s very easy to trigger this vulnerability, and it can double spend all assets on NEO.

\section{double-spend Attacks caused by Inconsistent Execution of Virtual Machine}
The consensus mechanism of smart contract platform essentially applies all transactions [$Tx_1$, $Tx_2$,...,$Tx_n$] to the initial $State_0$ in order, so that the entire network remains in the same state. During state machine replication, we require $State_n \times Tx_{n+1} \rightarrow State_{n+1}$ to be decisive. Essentially, $State_n \times Tx_{n+1} \rightarrow State_{n+1}$ is the execution process of the smart contract virtual machine to $Tx_{n+1}$. If there is a design or implementation vulnerability in the smart contract virtual machine, the virtual machine may have inconsistent execution(for the same input $State_n$ and $Tx_{n+1}$, the output $State_{n+1}$ is inconsistent), thus the attacker can use this problem to generate forks and double-spend attacks in the network. Below we will list multiple inconsistence vulnerabilities in the virtual machine we found on EOS and NEO and analyze the root causes.

\subsection{Remote code execution}
Previously, we published the report "EOS Node Remote Code Execution Vulnerability --- EOS WASM Contract Function Table Array Out of Bound"~\cite{RCE}. In this article, we found a buffer out-of-bounds write vulnerability in EOS. The exploit we wrote can successfully exploit this vulnerability to enable the EOS virtual machine to execute arbitrary instructions, thus fully controlling all EOS super node and verification nodes.

In essence, it is in the process of $State_n \times Tx_{n+1} \rightarrow State_{n+1}$, the attacker can escape the sandbox of the EOS virtual machine and result in arbitrary code execution, which can surely carry out the double-spend attack. The attack process is as follows:
     \begin{enumerate}
     \item The attacker constructs a malicious smart contract that can result in RCE and publish the contract to the EOS network.
     \item After the EOS super node parses the contract, it triggers the vulnerability and executes any instructions from the attacker.
     \item The attacker realizes a double-spend attack.
	\end{enumerate}
The vulnerability is very critical and is the first time that the smart contract platform has been attacked by remote code execution. You can read our previous report for details.

\subsection{Memory uninitialization vulnerability}
In the process of writing the exploit mentioned in "EOS Node Remote Code Execution Vulnerability --- EOS WASM Contract Function Table Array Out of Bound", we also exploited an undisclosed memory uninitialization vulnerability in EOS VM. In memory corruption attacks, memory uninitialization vulnerabilities can usually cause further problems such as information leakage, type confusion, etc., thus assisting attackers to bypass the mitigation measures of modern binary programs such as ASLR to further attack. However, in the smart contract virtual machine, there is a more direct exploit of memory uninitialized vulnerability, which can directly cause a double-spend attack. The following are detail of the memory uninitialized vulnerability that we used in the EOS RCE exploitation. This vulnerability can be directly used to double spend any token on EOS.

The WASM virtual machine uses the grow\_memory pseudo code to apply for new memory. In the initial implementation of EOS WASM grow\_memory, the requested memory was not reset. The content of the memory returned by grow\_memory is actually random. Then the attacker can construct a malicious contract to achieve a double-spend attack on any contract asset on EOS. The attack process is as follows:
     \begin{enumerate}
     \item The attacker constructs a malicious smart contract. A new memory address is obtained in the contract through grow\_memory.
     \item Read one bit of the address in the contract. [The bit may be 0 or 1 at this time, depending on the memory state of the running machine].
     \item The contract checks the content of that bit. If it is 1, then send the token to user A. If it is 0, send the token to user B, thereby achieving a double-spend attack.
	\end{enumerate}

\subsection{Memory uninitialization vulnerability}
In traditional memory corruption, the memory out of bounds read vulnerability will usually lead to information leakage, which will help us to bypass the mitigation measures of modern binary programs such as ASLR, and further exploit the system with other vulnerabilities. However, in the smart contract virtual machine, the memory out of bounds read vulnerability has a more direct use. It can directly cause double-spend attack. The following is an EOS memory out-of-bounds read vulnerability we found, which we can use to achieve a double-spend attack.

\begin{figure}
\centering
  \includegraphics[scale=0.5]{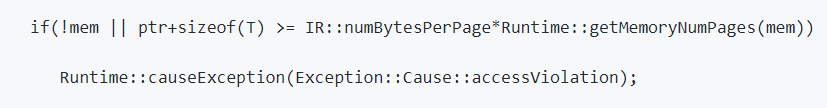}
    \caption{Offset into a WASM memory address}
    \label{EOS1}
\end{figure}

When EOS WASM converts an offset into a WASM memory address, the boundary check process is showed in Figure~\ref{EOS1}.

Here, the type of ptr is actually an I32 type, which can be a negative number. Then when:
\[-sizeof(T) < ptr < 0\]
ptr+sizeof(T) is a small number that can be pass this boundary check. Below this code, we can see the code:
\[T \&base = (T)(getMemoryBaseAddress(mem)+ptr)\]

The address of base will exceed the memory base address of the WASM, allowing the smart contract to implement memory out-of-bounds reading. [The contents of the memory address read depends on the current execution state of the virtual machine and can be considered random.] An attacker can exploit this vulnerability to implement a double-spend attack. The attack process is as follows:
     \begin{enumerate}
     \item The attacker constructs a malicious smart contract, and then uses the memory out of bounds read vulnerability to read one bit beyond the WASM memory base address. Here, this bit may be 0 or 1 depending on the state of the contract execution machine.
     \item The contract checks the content of the bit. If it is 1, then send the token to the user A. If it is 0, send the token to the user B; thereby achieving a double-spend attack.
	\end{enumerate}

\subsection{Inconsistent implementation of standard functions}
Summarizing the essence of the two examples of the double-spend attack above, the EOS contract actually reads the random variable in the execution process because of some memory vulnerabilities, thus breaking the consistency of the original virtual machine execution and causing a double-spend attack. In fact, the inconsistency of contract execution does not necessarily depend entirely on randomness. Here we will introduce a double-spend attack caused by inconsistencies in the implementation of standard C functions by various platforms (versions).

In the definition of C language standard, the return value of the memcmp function is required to be: less than 0, equal to 0, or greater than 0. However, in various implementations of the C language, the return value may be different (but still conform to the C standard). An attacker can take advantage of the inconsistency implementations of the C language standard to cause inconsistent execution in EOS virtual machines running on different systems, thereby achieving a double-spend attack. The attack process is as follows:

     \begin{enumerate}
     \item The attacker constructs a malicious smart contract, calls the memcmp function in the contract, and gets the return value.
     \item At this point, the return values of memcmp are different for different platforms and versions (even if the binary code of the EOS virtual machine is the same). The malicious contract checks the return value of memcmp and decides to transfer token to A or B. Thereby achieving the double-spend.
	\end{enumerate}
Here is the fix for this vulnerability in EOS VM in Figure~\ref{EOS2}:
 \begin{figure}
\centering
  \includegraphics[scale=0.5]{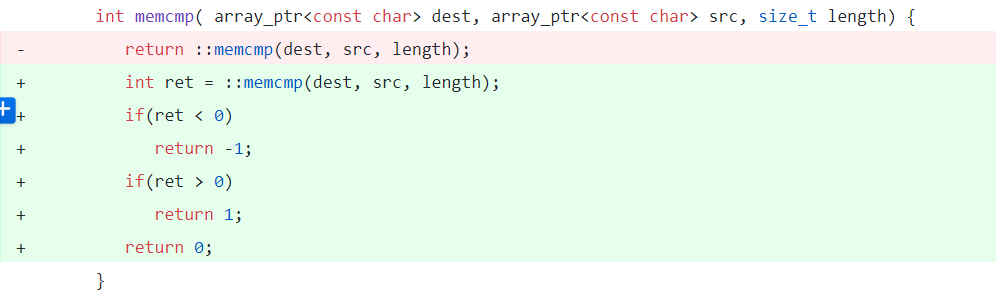}
    \caption{Fix of memcmp function in EOS VM.}
    \label{EOS2}
\end{figure}

EOS forces the return value of memcmp to be converted to 0, -1 or 1, thus preventing this inconsistent execution.

The problem with memcmp is caused by the inconsistency of the same standard implementation in the same programming language. In fact, the same blockchain project often has multiple implementations using different programming languages. Different programming language may also have difference in the same standard. For example, we found an inconsistent execution can be caused by the inconsistent implementation of the ECDSA standard using different programming language. The ECDSA signature standard requires that the private key x should not be zero. Such standards are strictly enforced in multiple cryptographic libraries written by Python and JS, but we have found that some of Golang's ECDSA libraries allow private key x=0 for signature generation and verification. Therefore, a malicious attacker can use this difference to implement a malicious contract for constructing inconsistencies in different implantation of the virtual machine (e.g.: Golang implementation and python implementation) of the same blockchain platform, thereby further achieve the double-spend attack.

\subsection{Inconsistent implementations of different versions}
The same blockchain project often has multiple implementations using different programming language. Implementations of different programming languages also have the potential for inconsistent execution. The above ECDSA case is an example. Big integer arithmetic is also a common example. For instance, in the implementation of NEO's C\# version and Python version, the BigInteger division operation can lead to inconsistent execution among different programming language implementations, resulting in a double-spend attack. Similar phenomenon has occurred in multiple blockchain projects.

\subsection{Other inconsistency issues}
Factors such as system time, random number, and floating point calculation are also causes of inconsistent execution of virtual machines. However, in our audit, we did not find such vulnerabilities in the popular public blockchain projects. Most blockchain projects are designed with these obvious issues in mind.

But the factors that may cause inconsistent execution may far exceed those we found above. In fact, some subjective factors (depending on the current operating state of the machine, which we call subjective factors) can cause inconsistent execution of virtual machines. For example, in 4G memory, 8G memory machines have different subjective boundaries for memory overflow (OOM) during execution. Attackers using OOM may cause inconsistent execution of virtual machines.

\section{double-spend Attacks Caused by Consensus Mechanism}
The double-spend attack caused by the consensus mechanism is actually a problem that has been fully discussed in the industry. However, various public blockchain schemes may still have a fork problem in the implementation of the consensus mechanism, resulting in a double-spend attack. In this section, we will give some example of double-spend attack caused by the consensus mechanism.

\subsection{VRF bypass in the ONT vBFT consensus mechanism}
Long range attack~\cite{Long} is a forking attack currently faced by all PoS consensus mechanisms. The attacker can choose not to fork the existing chain, but actually return to a chain state a long time ago (the attacker once occupied a large amount of money in this state), making a new chain with a longer jump to make the network mistakenly think it is the main chain, thus achieving double-spend attack. At present, there is no fundamental solution for the long range attack in the industry. We can only guarantee that the fork will not occur if the "Weak Subjectivity" does not occur.

The vBFT consensus mechanism in ONT proposes a method that relies on verifiable random functions (VRF)~\cite{VRF} to prevent malicious fork. The network first selects a set of block producers, block validators based on the VRF in the consensus network, and then completes the consensus by these selected nodes. Since the priority of each candidate block is determined by the VRF, it is difficult for the attacker to maintain his high priority for the maliciously forked blocks (if the attacker does not control the majority of the shares), so malicious forked chain will soon die. This giving vBFT a fast finality.

However, we found a vulnerability in the VRF in vBFT, which caused the user with the zero private key can generate the same VRF value for any block data. Specifically, VRF in vBFT is an implementation of the draft VRF standard proposed by Boston University~\cite{VRF2}. Specifically in Sections 5.1 and 5.2 of the draft, we can see the algorithm for ECVRF\_prove in Figure~\ref{ONT}:

 \begin{figure}
\centering
  \includegraphics[scale=0.4]{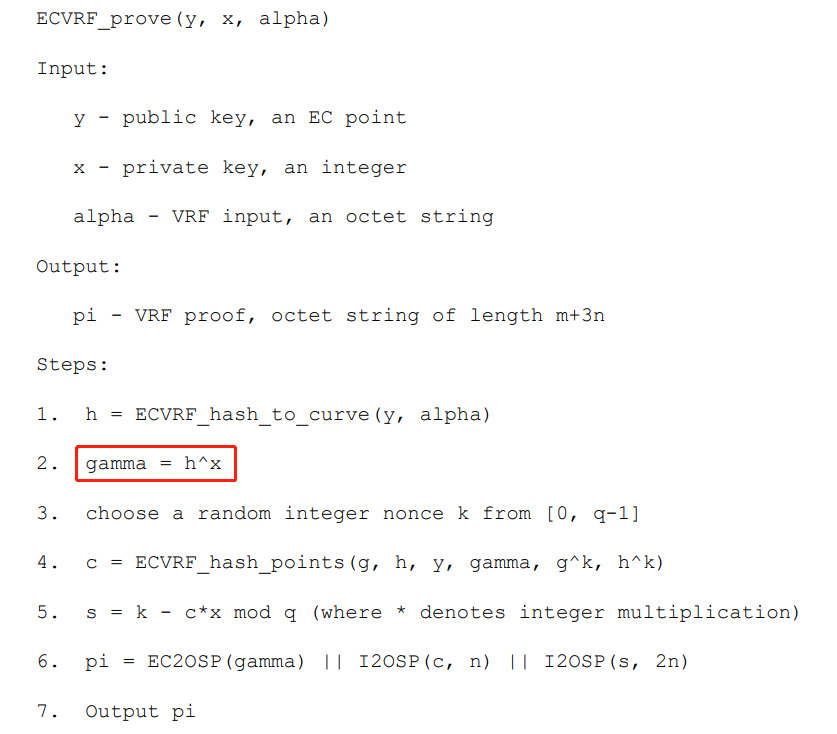}
    \caption{ECVRF\_prove in VRF standard draft.}
    \label{ONT}
\end{figure}

When x=0, it is still a legal private key in ONT’s implementation, and it can pass the ValidatePublicKey function. Then gamma will be a fixed point on the elliptic curve (infinity point). That is, for any input alpha, the value generated by the VRF is a fixed value. There is no randomness at all. This vulnerability can lead to an attacker using a fixed VRF value to control the “randomness” of the consensus algorithm, thereby controlling block generation for a long time.

\subsection{Fork in dBFT consensus mechanism}
The dBFT consensus mechanism~\cite{dBFT} can be considered as a POS+pBFT~\cite{pBFT} solution to achieve consensus in the entire network. In the original NEO and ONT’s implementation of dBFT consensus mechanism, we found that NEO and ONT have a fork problem. A malicious consensus node can generate a forked block, thus causing a double-spend attack. For details, please refer to our previous article: "Analysis and Improvement of NEO's dBFT Consensus Mechanism"~\cite{attackdBFT}.

\section{Mitigation Measure for double-spend Attack Caused by VM Inconsistent Execution}
To solve the double-spend attacks caused by the insufficient verification and flaw in consensus mechanism, it is necessary to dig into the specific detail in implementation. For the double-spend attacks caused by inconsistent execution in virtual machines, here we propose a general mitigation.

A simple way to solve the double-spend problem caused by inconsistent execution of the virtual machine is that the block producer hashes the global state $State_{n+1}$ after running the transactions, and then include the hash into the block. After receiving the block, the ordinary node compares the hash of the local state $State_{n+1}’$ with the hash of $State_{n+1}$. If they are equal, then no forks are generated. However, since the local data is growing first, the overhead of hashing the global state is enormous. In response to this problem, Ethereum~\cite{ethereum} used MekleTree structure to improve the performance while dealing with fork rollback issues. However, the Ethereum solution does not apply to blockchain projects that use other data structures to store state information. Here we propose an efficient solution which suitable for any data structures, the workflow is as follows:

     \begin{enumerate}
     \item In the block generating phase, the block producer records the write sequence [$write_db_1$,$write_db_2$,...,$write_db_n$] of the database in all the transaction runs in the block, and calculates the hash value $write_db_hash$ of the sequence.
     \item After the ordinary node receives the new block, it will verify the block first and then execute the transaction in the virtual machine. At the same time, the write sequence of these transactions to the database [$write_db_1'$,$write_db_2'$,...,$write_db_n'$] is recorded locally, and then $write_db_hash'$ is calculated. Check if it is equal to $write_db_hash$. If they are equal, then no inconsistent execution is considered to have occurred. If not, refuse to commit the sequence of write operations.
	\end{enumerate}

The core idea of this method is that the inconsistency of the virtual machine execution is due to the fact that various functional functions in the contract and the supporting of Turing completeness may introduce various uncertain factors, thus resulting in inconsistent execution. There are variety of complicated small reasons may lead to inconsistency. But let's take a step back. The essence of the double-spend attack is to modify the global state $State_{n+1}$, which is essentially a series of simple write operations (simple write operations often do not produce ambiguity). To prevent double-spend, you only need to match and verify all the write sequences. The overhead of matching and recording these write operations locally is very small, and recording the sequence of these write operations locally can also use for other issues such as fork rollback.

\section{Conclusion}
In this article, we introduced multiple cases of double-spend attack vulnerabilities we discovered on EOS, NEO and other large public blockchain platforms. We summarize various reasons for causing double-spend attacks, and propose a general mitigation measure. From the above analysis, we can see that the currently many blockchain projects still faces many security challenges. Various large blockchain projects are vulnerable to double-spend attacks.

\section{Acknowledge}
All of the vulnerabilities mentioned in this article have been fixed. In the process of solving those vulnerabilities, we found that EOS and NEO’s team are professional and efficient in responding to security issues. The security of those projects are also improved step by step. In a few months’ research on blockchain security, we have received more than \$300,000 award from various projects for reporting security bugs. We would like to say thanks to all the vendors.



\end{document}